\documentstyle[12pt]{article}
\author{A.Burinskii\\
Gravity Research Group, IBRAE Academy of Sciences\\
B.Tulskaya 52,Moscow 113191,Russia, e-mail:grg@ibrae.ac.ru}

\title{ Kerr Spinning Particle, Strings, and \\
Superparticle Models }
\begin{document}

\maketitle

\begin{abstract}
A combined model of the Kerr spinning particle and  superparticle is
considered. The structure of the Kerr geometry is presented in a complex form
as being created by a complex source.
A natural supergeneralization of this construction is obtained
corresponding to a complex "supersource".
Peforming a supershift to the Kerr and Kerr-Sen solutions we obtain
metrics of supergravity black holes with a nonlinear realization of
broken supersymmetry.

\end{abstract}
\par
\pagebreak
{\bf1.Introduction}
\par
\medskip
It was mentioned about 30 years ago  that the Kerr geometry displays
some remarkable features suggesting certain relationships with the spinning
elementary particles. In particular, the gyromagnetic
ratio of the Kerr-Newman solution is the same as that of the Dirac
electron. This fact stimulated treatment of the  models of spinning
particles based on
the Kerr-Newman geometry [1-6]. There were obtained some string-like
structures in the Kerr geometry. The first one is connected with a singular
ring of the Kerr solution [6-7]. Two others are linked with a complex
representation of the Kerr geometry (initiated by Lind and Newman [8])
in which the Kerr-Newman solution is considered as a retarded-time field
generated by a mysterious complex source propagating along a complex world
line. It was mentioned [9] that the complex world line is really a world
sheet or a special
type of string. The stringy boundary conditions for this complex
world line are connected with a third stringy structure of the Kerr
geometry - an orbifold [9].
\par
        A new and a very important period was started by
Witten (1992) who pointed out the role of black holes in string theory,
and also especially with the paper by Sen [10] who gave a generalization
of the Kerr solution to low energy string theory.
\par
It was shown [7] that near the Kerr singular ring the Kerr-Sen solution
acquires a metric similar to the field around a heterotic string.
Recently, much attention has also been paid to multidimensional Kerr
solutions and to a treatment of black holes as fundamental string states
[10,11] leading to a conclusion, suggested from different points of view,
that some black holes should be treated as elementary particles [12].
\par
    On the other hand, after obtaining supersymmetry,  great attention
has been paid to the models of spinning particles based on the Grassmann
anticommuting parameters ( D.V. Volkov and V.P.Akulov [13],
R.Casalbuoni [14], Brink and Schwarz [15], and others [16,17]), which has
also found an important application in superstring theory.
\par
    In this paper we consider one very natural way to combine the Kerr
spinning particle and superparticle models in such a manner that the
superparticle plays the role of a "source" of the supergeneralized
Kerr geometry.
Our treatment is based on the formalism by Debney, Kerr and Schild [2]
adopted to the above mentioned complex representation of the Kerr
geometry.
\par
The main idea of this work is extremely simple: to replace the mysterious
complex source of the Kerr geometry by a complex supersource which
can be obtained by an extra supershift.
\par
 Tugai and Zheltukhin [17] have recently shown that the application of
the supershift to Coulomb solution in a flat space gives rise to a Maxwell
supermultiplet of fields.
 On the other hand it was shown by Appel as early as in 1887 that complex
shift yields a ring-like singularity and specific Kerr's twofoldedness
 of space. Therefore, on the basis of these examples one can mention
that the methods of a complex shift and supershift have much in common
mathematically, though they lead to very different physical consequences.
In this paper examples of the simultaneous application of both above
transformations to the Kerr and to the Kerr-Sen solutions are given.
\footnote{In spite of a quite long story of supersymmetry the
number of known nontrivial supersolutions in electrodynamics and
supergravity is very small. Nontrivial supergravity solutions cannot
be obtained by a supergauge transformation from the corresponding known
solutions of Einstein's gravity. The only nontrivial super-BH solution
known to us is a supergeneralization of Reissner-Nordstr\"om
solution given by Aichelburg and G\"uven [20].}
As a result we derive the metrics of  rotating super black holes with
broken four-dimensional supersymmetry generated by a superparticle
source.
\par
\medskip
{\bf 2. Complex structure of the Kerr solution.}
\par
\medskip
Starting from the Kerr-Schild form of metric $ g_{ik} = \eta
_{ik} + 2h k_{i}k_{k};$
where $\eta _{ik}$=diag(-1,1,1,1) is the
auxiliary Minkowski metric in Cartesian coordinates
$(t,x,y,z)$, one can see that the main peculiarities of the Kerr solution are
connected with a form of the harmonic scalar function $h$ and vector field $k$
of principal null directions (PN congruence).
The function $h$ is the Appel potential
 \begin{equation} h= m Re(1/\tilde r),
\label{h} \end{equation}
where $\tilde r$ may be expressed in the oblate spheroidal coordinates
$r, \theta$ as
\par
\begin{equation} \tilde r = r + ia \cos \theta .\label{cdist} \end{equation}
It has a ring-like singularity at $r=\cos\theta=0$ which is a branch line of
the Kerr geometry. The space is covered by two sheets corresponding to
the positive and negative values of $r$.
The function $\tilde r$ may also be represented as a complex radial distance
\begin{equation} \tilde r =\sqrt { ( x_a-x_{0a})(x^a-x_0^a)},\quad a=1,2,3,
  \label{rcom}\end{equation}
from the complex point $x_0 =(0,0,ia)$. It involves a complex interpretation
of the Kerr solution, initiated by Lind and Newman [8], in
which the Kerr geometry is represented as a
retarded-time field generated by  a "complex point source" which propagates
in the complex Minkowski space $CM^4$
along a complex "world line"
$x^{i}_{o}(\tau),\quad (i=0,1,2,3),$
 parametrized by a complex time parameter
$\tau = t+i\sigma  = x^{o}_{o}(\tau) .$
This interpretation is also suggested by the analysis of the
field of principal null directions $k$ which is geodesic and shear free.
\par
An important role in this construction
 is played by  complex light cones,
 whose apexes lie  on  the  complex "world line" $x_{o}(\tau)$.
The complex light cone
\begin{equation}
 { K }= \{x: x =
x^{i}_{o}(\tau) + \psi ^{\alpha}_{R} \sigma ^{i}_{\alpha \dot {\alpha}}
\bar{\psi}^{\dot{\alpha}}_{L}  \label{psi} \}
\end{equation}
 may be split  into
two families of null planes: "right" $( \psi _{R}$ =const; $\bar{\psi
}_{L}$ -var.) and "left"$ ( \bar{\psi }_{L}$ =const; $\psi _{R}$ -var.).
The rays of the P.N. congruence $k(x)$ of the Kerr geometry  are
the tracks of these complex null planes (right or left) on the real  slice
of Minkowski space [6,9,21]. PN congruence propagates from a "negative"
sheet of 3-space
onto "positive" one crossing the disk spanned by the Kerr singular ring.
In the null coordinates $ u=(z+t)/ \sqrt{2};\quad
v=(z-t)/\sqrt{2}; \quad \xi =(x+iy)/\sqrt{2};
\quad \bar \xi=(x-iy)/\sqrt{2}$
we have
\begin{equation} k= k_{i} d x^i =
P^{-1}( du +\bar Y d\xi + Y d \bar \xi - Y \bar Y dv ),
\label{2.2} \end{equation}
where  $Y(x)$ is a complex projective spinor field
$Y={\bar \psi}^{\dot 2},\quad{\bar \psi}^{\dot 1}=1$.
\footnote{Here we use  the spinor notations of the book [19].}
\par
    The condition for a complex light cone to have a real slice is
\begin{equation}[x - x_0(\tau)]^2=0, \label{LC} \end{equation}
where $x$ is a real point. In the rest frame and with gauge $x_0^0 =\tau$
this equation may be split  as a complex retarded-time equation
\begin{equation}t-\tau =  \tilde r = - (x_i-x_{0i}){\dot x}_0^i.
\label{split}\end{equation}
It fixes the relation $ Im \tau = \sigma = a \cos (\theta)$
between the imaginary part of the complex time and a family of the null rays
with polar direction $\theta,\phi$.
\par
The Kerr theorem [2,6,21] allows one to describe the geodesic and shear-free
PN congruences in twistor terms via function $Y(x)$ which is a solution
of the equation\footnote{ The three parameters $ Y,\quad
\lambda _{1} = u + Y \bar \xi ,\quad \lambda _{2} = \xi - Y v  $
are  projective twistor coordinates.}
$ F(Y, \lambda
_{1},\lambda _{2}) = 0, $
  $F$ being  an analytical  function.
The complex radial distance $\tilde r$ may be expressed as
$ \tilde r = - dF/dY $.
Singular regions are defined as caustics of the congruence
satisfying the system of equations $F=0; \quad dF/dY =0.$
\par
For the Kerr congruence the function $F$ can be
expressed via parameters of the complex world line $x_{o}(\tau)$ [6,9,21]
\begin{equation} F \equiv (\lambda_1 - \lambda_1^0) K \lambda_2^0 -
(\lambda_2 -\lambda_2^0) K \lambda_1^0,\label{lam}\end{equation}
where $ K = [\partial_\tau x_0^i (\tau)] \partial_i, $
and $ \lambda_1^0, \lambda_2^0 $ are values of the twistor coordinates on
the world line $x_0(\tau)$. The resulting function $F$ is quadratic in $Y$
and the solution $Y(x)$ may be given in explicit form.
\par
Therefore, the Kerr solution may be represented as a retarded-time
field created by a mysterious "complex point source" propagating in the
auxiliary complex Minkowski space $CM^4$.
\par
\medskip
{\bf 3. Geometry generated by the super world line.}
\medskip
\par
Now we would like to generalize this complex retarded-time construction
to the case of complex "supersource" propagating along a super world line
\begin{equation}
X_0^i (\tau) = x_0^i(\tau) - i \theta\sigma^i \bar \zeta
 + i \zeta\sigma^i \bar \theta;
\quad \zeta^\alpha (\tau),\quad {\bar\zeta}^{\dot\alpha}(\tau).\label{SWL}
\end{equation}
Similarly to the above "real slice" we  introduce a
"B-slice" as a "body" of superspace [20], where the nilpotent part of
$x^i$ is equal to zero. The "real slice" is a real subset of the "B-slice".
The real slice condition (\ref{LC}) takes now the
form $ s^2=[x_i - X_{0i}(\tau)] [x^i - X_{0}^i (\tau)] = 0 $.
Selecting the nilpotent parts of this equation we obtain
 the above real slice condition (\ref{LC}) and the B-slice conditions
\begin{equation} [x^i-x_0^i (\tau)]
( \theta\sigma_i \bar \zeta
 - \zeta\sigma_i \bar \theta)=0; \label{odd1}\end{equation}
\begin{equation}
( \theta\sigma \bar \zeta
 - \zeta\sigma \bar \theta)^2 =0.\label{odd2}\end{equation}
The equation (\ref{odd1}) may be rewritten using (\ref{psi}) in the
form
\begin{equation}
  (\theta^\alpha\sigma_{i\alpha\dot\alpha}\bar\zeta^{\dot\alpha}
 - \zeta^\alpha\sigma_{i\alpha\dot\alpha}\bar\theta^{\dot\alpha})
\psi^\beta \sigma^i_{\beta\dot\beta} {\bar\psi}^{\dot\beta}
=0\label{odd4}\end{equation}
which yields
\begin{equation}
\bar\psi \bar\theta =0,\qquad\bar\psi \bar\zeta =0,
\label{odd5}\end{equation}
which in turn is a condition of proportionality of the commuting spinors
$\bar\psi(x)$ and anticommuting spinors $ \bar\theta$ and $\bar\zeta$
providing the left  null superplanes to reach the B-slice.
Taking into account that ${\bar \psi}^{\dot 2}=Y (x),
\quad{\bar \psi} ^{\dot 1}=1$ we obtain
\begin{equation}
{\bar\theta}^{\dot 2} = Y (x){\bar\theta}^{\dot 1} ,\quad
{\bar\theta}^{\dot \alpha} = {\bar\theta}^{\dot 1}
{\bar\psi}^{\dot\alpha},\quad
{\bar\zeta}^{\dot 2} = Y (x){\bar\zeta}^{\dot 1} ,\quad
{\bar\zeta}^{\dot \alpha} = {\bar\zeta}^{\dot 1} {\bar\psi}^{\dot\alpha}.
\label{SRS}
\end{equation}
\par
It also gives that $ \bar\theta \bar\theta= \bar\zeta \bar\zeta=0, $
and equation (\ref{odd2}) is satisfied automatically.
\par
Therefore, the B-slice condition fixes a correspondence between the
coordinates  $ \bar\theta, \quad \bar\zeta$ and twistor null planes forming
the Kerr congruence, and consequently the coordinate $\bar\zeta$ of the
super world line must be engaged partially to provide  B-slice and
parametrize the "left" complex planes and the null rays of the Kerr
congruence. The conjugate sector also gives
${\theta}^{\alpha} = \bar Y (x){\theta}^{1}$; however,
the coordinate of the super world line $\zeta$ remains independent,
and can be left as an arbitrary function of time.
\footnote{The coordinates
$\theta^1$, ${\bar\theta}^{\dot 1}$, and ${\bar \zeta}^{\dot 1} $
are independent too.}
Therefore, the roles of the chiral and antichiral
Grassmann coordinates of the super world line are to be divided.
\par
       The retarded time equation (\ref{split}) takes now the form
$ t- T =  \tilde R = \tilde r + \eta, $
where $\tilde R = - (x_i-X_{0i}){\dot X}_0^i $ is a superdistance.
The "body-part" of this equation satisfies the above relation (\ref{split}),
$T=\tau - \eta $ is a supertime containing the nilpotent term
\begin{equation}
\eta = i \theta\sigma^0 \bar \zeta -i \zeta (\tau)\sigma^0\bar \theta.
\label{eta}\end{equation}
In the stationary case $\dot x_0^i =(1,0,0,0)$, $\dot\zeta=0$  on the
B-slice $\tilde R$ takes the simple form
\begin{equation}\tilde R= r + i a \cos \theta
 + i \theta\sigma^0\bar\zeta -i \zeta(\tau)\sigma^0\bar\theta .
\label{dist}\end{equation}
  The corresponding supergeneralization of the Kerr theorem may be achieved
by substitution of the super world line $X_0(\tau)$ instead of $x_0(\tau)$
in the function $F$. As a result one can obtain a superfield $Y(x)$ which
on the B-slice takes the usual form since all the nilpotent terms
disappear.
From the Kerr theorem one  obtains  the general expression for superdistance
out of B-slice
$$\tilde R =- d \hat F /d{ Y} =  \tilde r -
i[x^i-x_0 ^i(\tau)] \dot\zeta(\tau) \sigma_i  {\bar \theta} -
i[{\dot x}_0 ^i(\tau) + i \dot\zeta (\tau)\sigma^i {\bar\theta}]
( \theta\sigma_i{\bar\zeta}  - \zeta \sigma_i\bar\theta), $$
which may be useful when applying the (anti)chiral differential
operators $D_{\alpha}, \quad\bar D_{\dot\alpha}$ [17,19].
\par
\medskip
{\bf 4. Supershift of the Kerr solution.}
\par
\medskip
One can note that the Kerr solution is a particular solution
of supergravity with vanishing spin-3/2 field, and
that in the stationary case $\dot X_0=(1,0,0,0), \quad \dot \zeta=0$
the solution with supersource (\ref{SWL}) can be obtained from the Kerr
solution by a supershift
\begin{equation}
x^{\prime i}  = x^i + i \theta\sigma^i \bar \zeta
 - i \zeta\sigma^i \bar \theta;
\qquad
\theta^{\prime}=\theta + \zeta ,\quad
{\bar\theta}^{\prime}=\bar\theta + \bar\zeta, \label{SG}
\end{equation}
which is a "trivial" supergauge transformation in supergravity.
However, the subsequent imposition of a B-slice constraint is a nonlinear
operation breaking four-dimensional supersymmetry [13,18,19]. As a result
the arising spin-3/2 field cannot be gauged away.
\par
Starting from tetrad form of the Kerr solution  $ds^2=2e^1 e^2 +2e^3 e^4,$
where \begin{equation}
e^1 = d \xi - Y d v= \partial_{\bar Y} \psi \sigma_i
\bar \psi dx^i/\sqrt{2};
\quad e^2 = d \bar \xi - \bar Y d v =
 \partial_{ Y} \psi \sigma_i \bar \psi dx^i/\sqrt{2};
\label{KS1}
\end{equation}
\begin{equation}
e^3 =du+ \bar Y d \xi  + Y d \bar \xi - Y \bar Y d v=
\psi \sigma_i \bar \psi dx^i/\sqrt{2};
\label{KS3}
\end{equation}
\begin{equation}
e^4 = - \partial_{\bar Y} \partial_Y e^3 - h e^3,\label{KS4}
\end{equation}
and using the expressions
\begin{equation}
d x^{\prime i} = d x^i + i(\theta^1 \bar \zeta^{\dot 1})(\partial_{\bar Y}
\psi\sigma^i\bar\zeta)d\bar Y  +
 i(\theta^1 \bar \zeta^{\dot 1})
(\psi \sigma^i \partial_Y \bar\psi) dY  +
i \bar \theta^{\dot 1}(\zeta \sigma^i \partial_Y \bar\psi)dY,
\end{equation}
obtained from the coordinate transformations (\ref{SG}) under constraints
(\ref{SRS}), and also substitution $\tilde R \rightarrow \tilde r$, one
obtains the following tetrad
\begin{equation}
e^{\prime 1} = e^1 + ( A - C^1 {\bar \theta}^{\dot 1} ) dY,\qquad
e^{\prime 2} = e^2 + Ad\bar Y ,
\label{S12}
\end{equation}
\begin{equation}
e^{\prime 3} = e^3 - C^3 {\bar \theta}^{\dot 1}dY,\qquad
e^{\prime 4} = dv + \tilde h e^{\prime 3} ,
\label{S34}
\end{equation}
where $dY={\tilde R}^{-1}(P e^1 -P_{\bar Y} e^3), $
and
\begin{equation}
 A=i\sqrt{2}(\theta^1 {\bar \zeta}^{\dot 1}) ,\qquad
C^a =ie^a_i(\zeta \sigma^i \partial_Y \bar \psi),
\label{AC}
\end{equation}
\begin{equation}
 \tilde h= m(Re{\tilde R}^{-1})/P^3 , \qquad P={\sqrt 2}^{-1}(1+Y \bar Y).
\label{hY}
\end{equation}
As a result we obtain the metric of a super black hole $ds^2=e^{\prime 1}
e^{\prime 2} +e^{\prime 3} e^{\prime 4}$ with broken four-dimensional
supersymmetry.
 For parameters of spinning particles it corresponds
to a specific state of a "black hole"  without horizons and very far
from extreme.
\par
This derivation of a super-Kerr metric is similar to the first
derivation of  the Kerr-Newman solution by complex shift from the
Reissner-Nordstr\"om metric given by Newman and collaborators (1965).
The first use of a complex shift in scalar electrodynamics is traced back
to Appel (1887) who discovered the  potential $e Re (1/{\tilde r})$
characterized by a typical Kerr's singularity and twofoldedness of space.
The first use of supershift in electrodynamics was considered in the recent
work by Tugai and Zheltukhin [17].
        As a result a supermultiplet of Maxwell fields
was generated from the Coulomb solution.
\par
Therefore, at the moment there are several known applications of the
method in consideration.
For example, the simplest interesting new solutions can be obtained by
simultaneously performing the complex shift and supershift to the Coulomb
solution in flat space. Similarly, a supergeneralization of the Kerr-Newman
solution leading to a supermultiplet of Maxwell fields on the Kerr
background may be obtained, as well as a supergeneralization of the
Kerr-Sen solution.
\par
\medskip
{\bf 5. Supershift of the Kerr-Sen solution to dilaton-axion gravity}
\par
\medskip
The Kerr-Sen solution, a generalization of the Kerr solution to
low energy string theory [10], may be written in the  form [7]
 \begin{equation} ds^2_{dil}=2e^{-2(\Phi - {\Phi}_0)} {\tilde e}^1
{\tilde e}^2 +
2{\tilde e}^3 {\tilde e}^4, \label{dil} \end{equation} where
\begin{equation}
{\tilde e}^1 =(PZ) ^{-1} dY ,
\qquad{\tilde e}^2 =(P\bar Z) ^{-1} d \bar Y ,
 \label{te12}\end{equation}
 \begin{equation}
{\tilde e}^3 = P^{-1} e^3 ,
 \label{te3}\end{equation}
 \begin{equation}
{\tilde e}^4 = dr + iaP^{-2}(\bar Y dY - Y d \bar Y) +
 (H_{dil} -1/2)e^3,
 \label{te4}\end{equation}
and  \footnote{The definition of $H_{dil}$ in [7] is different and
contains an extra factor of 2.}

 \begin{equation}
H_{dil}  = M r/ \Sigma_{dil};\quad
\Sigma_{dil}=e^{-2(\Phi - {\Phi}_0)}(Z{\bar Z})^{-1};
 \label{H}\end{equation}
 \begin{equation}
e^{-2(\Phi - {\Phi}_0)} = 1 + (Q^2/2M)(Z+{\bar Z});
\qquad Z^{-1} \equiv \tilde r.
 \label{dz}\end{equation}
The field of principal null directions is $\tilde e^3$.
Following eq.(6.1) of [2] this tetrad is related to the Kerr-Schild
tetrad (\ref{KS1}),(\ref{KS3}),(\ref{KS4}) as follows
 \begin{equation}
{\tilde e}^1 =e^1 - P^{-1} P_{\bar Y} e^3, \qquad
{\tilde e}^2 =e^2 - P^{-1} P_{Y} e^3,
 \label{12}\end{equation}
 \begin{equation}
 {\tilde e}^3= P^{-1} e^3,
 \label{3}\end{equation}
 \begin{equation}
{\tilde e}^4 = Pe^4_{dil} + P_{ Y} e^1 + P_{\bar Y} e^2
-P_{Y}P_{\bar Y}P^{-1}e^3.
 \label{4}\end{equation}
Therefore the Kerr-Sen metric (\ref{dil}) may be reexpressed
in a form containing the Kerr-Schild tetrad $e^a$,  dilaton factor
$e^{-2(\Phi - {\Phi}_0)} $, and a deformed function
\begin{equation}
 H_{dil} = h e^{2(\Phi - {\Phi}_0)}
\end{equation}
instead of the function $h$ in the tetrad vector $e^4$ given by (\ref{KS4}).
\par
It was shown in [7] that the field of principal null directions $ e^3$
survives in the Kerr-Sen solution and retains the property of to being
geodesic
and shear free. It means that the Kerr theorem is applicable to this
solution too, as well as the above geometrical construction if tetrad
is expressed in Cartesians coordinates $x,y,z,t$.
The corresponding "supershifted" solution is obtained by
the substitution  $\tilde R \rightarrow \tilde r$ in the expression
for the dilaton factor  and  by using
the "supershifted" Kerr-Schild tetrad (\ref{S12}),(\ref{S34}),(\ref{AC})
in the expressions (\ref{12}),(\ref{3}),(\ref{4}).
\par
Summarizing, we find that metric is given by
 \begin{eqnarray}
 ds^2_{dil}=2e^{-2(\Phi - {\Phi}_0)} {e}^{\prime 1}
{e}^{\prime 2} + 2{e}^{\prime 3} { e}^{\prime 4}_{dil}
+2[1-e^{-2(\Phi - {\Phi}_0)}] ( P_{Y} e^{\prime 1}\nonumber\\
+ P_{\bar Y} e^{\prime 2}
- P_{Y}P_{\bar Y}P^{-1}e^{\prime 3}) e^{\prime 3}/P,
\label{pd1} \end{eqnarray}
where
\begin{equation}
e^{\prime 1} = e^1 + ( A - C^1 {\bar \theta}^{\dot 1} ) dY,\qquad
e^{\prime 2} = e^2 + Ad\bar Y ,
\label{SS12}
\end{equation}
\begin{equation}
e^{\prime 3} = e^3 - C^3 {\bar \theta}^{\dot 1}dY,\qquad
e^{\prime 4}_{dil} = dv + H_{dil} e^{\prime 3} ,
\label{SS34}
\end{equation}
and where $e^a $ are given by (\ref{KS1}),(\ref{KS3}),(\ref{KS4}),
and also
\begin{equation}
dY={\tilde R}^{-1}(P e^1 -P_{\bar Y} e^3),
\label{Y1}
\end{equation}
\begin{equation}
 A=i\sqrt{2}(\theta^1 {\bar \zeta}^{\dot 1}) ,\qquad
C^a =ie^a_i(\zeta \sigma^i \partial_Y \bar \psi),
\label{AC1}
\end{equation}
\begin{equation}
 H_{dil}= e^{2(\Phi - {\Phi}_0)} M Re {\tilde R}^{-1} /P^3 ;
\quad P=(1+Y \bar Y)/ \sqrt{2}, \label{HP}
\end{equation}
\begin{equation}
e^{-2(\Phi - {\Phi}_0)} = 1 + (Q^2/M) Re {\tilde R}^{-1}.
\label{sd}
\end{equation}
\par
\medskip
{\bf 6. Conclusion}
\par
\medskip
As we pointed out in Introduction the Kerr geometry contains string-like
structures and one of them is the Kerr singular ring. The gravitational field
near this ring is similar [7] to the field around a heterotic string [10].
In the super-Kerr geometry we find out some extra suggestions of this
relationship.
        In the presented superblackhole metrics the four-dimensional
supersymmetry is broken because of the nonlinear realization of
supersymmetry caused by B-slice constraints. However, there survives
(2,0)-supersymmetry based on the complex time parameter $\tau$ and
anticommuting superpartners  $\bar\theta^{\dot 1}$ and $\theta^{1}$.
It is known from the analysis of the Kerr theorem [9,21] that only an
analytic
dependence in the even function $x_0(\tau)$ is admissible. On the other
hand, during the above consideration we did not meet the demands for
the Grassmann parameter $\zeta(\tau)$ to be analytic in $\tau$.
It means that the arising (2,0)-superfields  can depend on
$\tau$ and $\bar\tau$ leading to  both  right and left
modes in the fermionic sector that must induce traveling waves along the
Kerr singular ring.
\par
It has also to be noted that for the known parameters of spinning particles
the angular momentum is very high regarding the mass parameter, and the
corresponding black holes are to be in a specific state "...which is neither
`black' and nor `hole'..." [24]. In this case the ring-like
singularity is naked, and space is branched on two sheets,
$r>0$ and $r<0$ respectively. There appears a problem of the real source of
the
Kerr solution in addition to the mysterious complex supersource considered
above.
\par
 To avoid this twofoldedness the 'negative' sheet of space is truncated
and a matter source is placed on the disk $r=0$ spanned by the
Kerr singular ring. Such disk- or membrane-like sources of the Kerr
solution were considered in Einstein's gravity [3,5,22,23], as well
as in low energy string theory [7,11]. However, the analysis shows [3,5]
that very exotic properties of material are necessary to provide a continuity
of metric by  crossing the disk.
It was obtained that the Kerr disk has to be in a rigid relativistic
rotation and built of the material having a pseudovacuum character, zero
energy density in a corotating coordinate system.
\footnote{A more complete list of references to the problem of the Kerr source
can be found in [9].} There is no classical matter possessing this property,
and some attention was also paid to a possible role of quantum effects
[22].
\par
Since supersymmetry takes an intermediate position between the classical
and quantum regions one could expect that in some cases it could provide the
necessary pseudovacuum character of matter, especially taking into account
the remarkable cancellation contributions of fermionic and bosonic fields.
       It gives rise to a hope that the old problem of the source of the
Kerr solution could find its resolution in a composite source built of a
supermultiplet of matter fields.
\par
Obtaining the corresponding supersolutions describing the real
matter sources compatible with the Kerr supergeometry is a
very important problem in future investigations.
\par
        In conclusion I would like to thank A. Zheltukhin, V. Tugay,
V. Akulov and G. Alekseev for very useful discussions, I am also very
thankful to P. Aichelburg for useful discussions and hospitality at the
Vienna University and also to many participants of seminars in Dubna,
Kharkov, and Vienna for useful comments.
\par
\pagebreak
REFERENCES:
\par
\medskip
1. Carter B., Phys. Rev. {\bf 174}(1968) 1559.
\par
2. Debney G.C., Kerr R.P., Schild A., J.Math.Phys.,
{\bf 10}(1969) 1842.
\par
3. Israel W., Phys. Rev. ${\bf D2}$ (1970) 641.
\par
4. Burinskii A., Sov. Phys.  JETP {\bf 39}(1974) 193.
\par
5. L\`opez C.A., Phys. Rev. ${\bf D30}$ (1984) 313.
\par
6. Ivanenko D. and  Burinskii A., Izv. VUZ Fiz.
(Sov. Phys. J. (USA)).,nr.{\bf 7} (1978)113
\par
7. Burinskii A., Phys.Rev.D52(1995)5826; hep-th/9504139
\par
   Nishino H., Stationary axi-symmetric black holes, N=2 superstring, and
selfdual gauge of gravity fields, preprint UMDEPP 95-111; hep-th/9504142
\par
8. Lind R.W. and Newman E.T., J. Math. Phys.,{\bf 15}(1974)1103.
\par
9. Burinskii A., String - like Structures in Complex Kerr Geometry,
   Proc. of the Fourth Hungarian Relativity Workshop Edited by R.P. Kerr
and Z. Per\'jes, Academiai Kiado, Budapest 1994, Preprint gr-qc/9303003,
\par
   Burinskii A., Phys.Lett. {\bf A 185}(1994)441;
\par
10. Sen A., Phys.Rev.Lett., {\bf 69}(1992)1006,
\par
    Horowitz G.T. and Sen A., Phys. Rev. {\bf D 53} (1996)808
\par
11. Dabholkar A., Gauntlett J., Harvey J. and Waldram D., Nucl. Phys.
   {\bf B 474}(1996) 85.
\par
12. Sen A. Modern Phys. Lett. {\bf A 10}(1995)2081, Nucl.Phys {\bf B46}
(Proc.Suppl)(1996)198.
\par
  't Hooft G., Nucl. Phys. {\bf B335}(1990)138
\par
Holzhey C. and Wilczek F., Nucl. Phys. {\bf B380}(1992)447; hep-th/9202014
\par
13. Volkov D.V. and Akulov V.P. Pis'ma Zh. Eksp.Teor.Fiz. {\bf 16}621(1972)
\par
14. Casalbuoni R., Nuovo Cim., {\bf 33A},389,1976\par
\par
15. Brink I. and Schwarz J.,Phys.Lett.{\bf 100B}(1981)310 \par
\par
16. Pashnev A. and Volkov D., Teor.Mat. Fiz. {\bf 44} (1980)310.
\par
    Berezin F.A and Marinov M.S., Pis'ma ZhETP,{\bf 21}(1975)678 \par
\par
    Sorokin D.P., Tkach V.I.,  Volkov D.V. and Zheltukhin A.A.,
    Phys. Lett.{\bf B 216 }, 302 (1989)
\par
17. Tugai V. and Zheltikhin A. Phys.Rev. {\bf D51} (1955) R3997
\par
18. Deser S. and Zumino B. Phys. Rev. Lett., {\bf 38}(1977) 1433
\par
19. Wess J. and Bagger J. Supersymmetry and Supergravity, Princeton,
    New Jersey 1983
\par
20. Aichelburg P.C. and G\"uven R. Phys. Rev. Lett. {\bf 51} (1983) 1613
\par
21. Burinskii A., Kerr R.P. and Perjes Z., 1995, Nonstationary Kerr
    Congruences, Preprint gr-qc/9501012
\par
22. Burinskii A. Phys. Lett. {\bf B 216}(1989)123
\par
23. Burinskii A. Espec. Space Explorations, {\bf 9 (C2)}(1995) 60,
    Moscow, Belka, Preprint hep-th/9503094
\par
24. Townsend P.K. Supergravity Solitons and Non-Perturbative Superstrings.
    Proc. of 1995 Trieste Spring Superstring School and Workshop.
    Preprint hep-th/9510190
\end{document}